\documentstyle[aps,pre,epsf]{revtex}

\begin{document}

\twocolumn[\hsize\textwidth\columnwidth\hsize\csname
@twocolumnfalse\endcsname

\title{Stable propagation of an ordered array of cracks during directional drying}
\author{E. A. Jagla}
\address{Centro At\'omico Bariloche and Instituto Balseiro, 
Comisi\'on Nacional de Energ\'{\i}a At\'omica, (8400) Bariloche, Argentina}
\maketitle
\begin{abstract}
We study the appearance and evolution of an array of parallel 
cracks in a thin slab of material that is directionally dried, and show that 
the cracks penetrate the material uniformly
if the drying front is sufficiently sharp.
We also show that cracks have a tendency to become evenly spaced 
during the penetration.
The typical distance between cracks 
is mainly governed by the typical distance
of the pattern at the surface, and 
it is not modified during the penetration. 
Our results agree with recent experimental work, and can be 
extended to three dimensions to describe the properties 
of columnar polygonal patterns observed in some geological
formations.
\end{abstract}


\vskip2pc] \narrowtext

\section{Introduction}

Cracks appearing when a material shrinks are
common in everyday life. The most popular examples are probably 
the cracks appearing in
paint layers, and those in the surface of muds. 
In the case of paints, there is 
a superficial layer of material that shrinks on top of a substrate
to which it is attached. For muds, the superficial layer and the substrate
are the same material, but difference in the humidity concentration produces
the cracking at the surface.
In these two cases, the shrinking is due to changes in the humidity 
concentration within the sample, but it can also be due to the existence
of non-uniform temperature distributions, which produces stresses and 
generates cracking. The problem of surface fragmentation 
has been studied in the last
years, both theoretically\cite{theor} and experimentally\cite{exper}.

There are situations in which cracks appear at the surface 
of the material, and penetrate into the sample later on. A well known
example corresponds to the columnar fracturing of basaltic lava, 
seen in many different geographical locations\cite{basalts}.
A detailed description of this problem has only recently been 
foreshadowed\cite{jr0,jr1}, and still some points remain obscure.
The two dimensional equivalent of the columnar structure of basalts
corresponds to the case of a two dimensional material that dries
(or cools down) starting from a free edge. 
As in the three dimensional case, drying creates internal stresses 
in the material that generate cracks, first appearing at the free edge. 
As the drying front propagates to the interior the cracks also do so in turn.
There have been different experimental realizations of this phenomenon.
In one of them\cite{alan-lim} the material was a very thin layer of a colloidal dispersion
placed between two glasses, with one free edge, from which humidity can escape. 
In other experiments\cite{canad}, a slurry of Al$_2$O$_3$-water mixture 
was deposited onto a substrate, and
a glass was placed a few millimeters above it. A low, ultra dry $N_2$ gas breeze was
injected in the slot above the sample. The $N_2$ became saturated with water 
as it passed over the
material, and a rather sharp drying front propagates with time 
in the same direction than the air flow. A third realization corresponds
to the drying of thin films of aqueous silica sol-gel, where very nice patterns
have been observed\cite{hull}. In all these cases the propagation of a set
of parallel cracks has been observed.

A detailed theoretical description of this phenomenon is lacking, and
we will make an attempt in this direction for the two dimensional case. 
In the next section we will make a general description of the phenomenon, 
emphasizing the difference with a more standard fracture mechanics problem.
In Section III we present the calculations of elastic and crack energy
for the case in which the drying (or cooling) front is sharp. In Section IV
we discuss the conditions under which the crack front (formed by the
tips of all cracks) is stable, and progresses into the material when the
drying front advances.
In Section V we show that the array of cracks 
generated at the surface of the material becomes evenly spaced
when penetrates the sample, and that the typical width of the stripes is
determined mainly during the first stage of the process, near the surface.
In Section VI we briefly comment upon the effects of other drying
(or cooling) conditions than the one studied previously.
Section VII contains some implications of our work for the study of the
three dimensional case.
Finally, in Section VIII we summarize and conclude.

\section{Quasistatic fracture mechanics. Energetic and stress analysis}

One remarkable thing about the experiments cited in the previous 
section \cite{alan-lim,canad,hull} is the fact that a large number 
of cracks penetrate the sample in a coordinated and quasistatic manner, 
as the external conditions (the humidity profile) change. Then
this problem is qualitatively different of a typical problem in 
fracture mechanics, where the propagation of a crack is usually an
abrupt phenomenon and typically leads to the failure of the material.
In standard fracture mechanics, stress and energetic analysis
are two different ways of predicting the evolution of
the fracturing process\cite{broberg}. However, the equivalence of both
approaches is not clear\cite{setna}, particularly in cases in which
the cracks propagate at large speeds, which is almost always the case when
failure occurs.

The situation is different in our case. Cracks propagate
only because the external humidity profile changes. If the external conditions
were stationary, crack advance would be arrested. 
If we think on the configuration of the system as a point
in configuration space (the space spanned by all coordinates of all particles of 
the material), at each moment the system is at one minimum of the
energy landscape. As the humidity profiles changes the landscape changes itself,
the minimum on which the system is located shifts,
and the configuration of the system adapts so as to remain in the shifted
local minimum. This is what we understand as a quasistatic propagation
of cracks.\cite{nota10}

Under these conditions the propagation of cracks in a two dimensional
geometry can be studied by two
different but equivalent procedures:
Stress analysis consists on the calculation of the stress
intensity factors\cite{broberg} $K_I$ and $K_{II}$ of the opening and
shearing modes at the tips of 
the cracks present in the system. The propagation will occur 
in the direction along which $K_{II}=0$, which coincides with
that for which $K_I$ is maximum. Propagation actually occurs only if the
energy relieved by the advance of the cracks is enough to overcome
the fracture energy needed to elongate the cracks. In the quasistatic
case, these two energies will differ only infinitesimally. 

In the energetic procedure to calculate crack advance, the total energy 
(including elastic and crack energy) after virtual advances of the cracks
is calculated. Cracks will actually propagate 
only if this propagation reduces the total energy. Under quasistatic
advance the energy reduction during propagation is infinitesimally small,
and typically, there is only one possible direction of propagation for each crack,
all other propagation directions would produce an
increase of the total energy of the system.

Both approaches are equivalent in the case of quasistatic advance of the cracks.
We will use stress analysis or energetic analysis according to
convenience in each case.

\section{Estimations of elastic and crack energy}

We will consider an isotropic and homogeneous material, and assume that linear 
elasticity can be applied\cite{elasticidad}. 
It is well known  that in this case the material possesses
only two independent elastic constants\cite{landau}. We will take as 
these two parameters the bulk modulus 
$C$ and the Poisson ratio $\nu$. The expansion or contraction
properties of the material are described by a humidity
expansion coefficient $\alpha$, which is formally equivalent to the 
thermal expansion coefficient, i.e., the relative change of
linear size  $\delta L/L$ of a piece of material after changing 
the humidity concentration $h$ by some quantity $\Delta h$ is given by
\begin{equation}
\delta L/L=\alpha \Delta h.
\end{equation}

The ideal situation we will address consists of a semi-infinite 
two-dimensional sample that is being dried from its surface. 
The drying 
will be considered to be non-homogeneous, and we will model it 
by a humidity profile that is given by some 
function $h(z,t)$ depending on depth $z$ within the material, and time $t$. 
The precise form of this function
will be specified later for different experimental situations, but it is 
important to point out that we do not consider the case in which the 
appearance of cracks in the system modifies itself the drying process.
Then our analysis applies more to cases as that described in \cite{canad} and
possibly \cite{hull}, but much less
to that in \cite{alan-lim} where evaporation of humidity through the cracks 
seems to be relevant(see also \cite{bahr}). Although in the experimental situation the material 
usually lies on top of a substrate, we will study the case in which there 
is no interaction with this substrate. In the experiments of Ref. \cite{canad}
this is achieved by introducing a layer of some slippery material
between the sample and the substrate. Then, all the stresses on the sample
are originated internally, and are due to the existence of a non-uniform 
humidity concentration.

It is known that in an isolated (namely, not clamped) piece of 
homogeneous material placed under a constant thermal (or humidity) gradient
all stresses vanish in linear elastic approximation \cite{lineart}. Under these
conditions cracks cannot appear at all, or if already present from the beginning
its propagation is completely halted. It is in fact 
crucial for the propagation of cracks that the humidity gradient it is not constant
within the sample.

We will take the edge of the semi-infinite sample as
the $z=0$ line, and $z>0$ in the interior. For convenience, we will also
refer to the edge as the `surface', and consider it to be horizontally
placed, in such a way that the cracks propagate down the material.
At $t=0$ the whole material is supposed to have 
a constant humidity concentration $h_0$. In this situation the material is unstressed.
Humidity concentration is assumed to decrease with 
time $h(z,t>0)\le h_0$. For any reasonable 
experimental realization of the
drying process occurring from the $z=0$ free surface, it is clear that 
at any time, well inside the material we should reach the original 
humidity concentration, i.e., $h(z\rightarrow \infty,t)=h_0$. Then 
the majority of the material is always at constant humidity $h_0$. 
In this region the sample must be unstressed, otherwise it will 
store an infinite amount
of energy. Then a boundary condition for our problem will be that 
stresses go to zero as $z$ go to infinity.
As a simplification of a possible experimental situation,
we will consider the case of an abrupt 
drying profile (see Fig. \ref{f0}),
namely $h(z,t)=h_0$ for $z>z_0(t)$, and $h(z,t)=h_1$ 
for $z<z_0(t)$, with $h_1<h_0$, and $z_0(t)$ being an increasing function 
of time. 

A given set of cracks will always correspond to a local minimum
of the total energy of the system.
The total energy is the sum of two well different parts. One is the fracture
energy, namely, the energy spent in the creation of all fractures present in the
sample. This is typically proportional to the total length of cracks in
the material.
The second part is the elastic energy stored in the sample.
Let us suppose we have an evenly spaced
array of cracks, defining stripes of width $l$, which have penetrated down to
some distance $\bar z$, with the drying front being located 
at some position $z_0$.

The elastic energy $e_{\rm el}$ stored in the material per unit
of horizontal length must be proportional to the bulk modulus of the 
material $C$, and to the second power of the typical change in linear
density caused by the humidity gradient (this change being $\alpha \Delta h$, with
$\Delta h\equiv (h_0-h_1)$). This is at the basis of linear elasticity
theory. Using the fact that $l$
and $\bar z-z_0$ are the only two relevant lengths for this geometrical
configuration\cite{zz},
dimensional analysis allows to 
write down the following expression for $e_{\rm el}$ 

\begin{equation}
e_{\rm el}=(\alpha \Delta h)^2 C l {g} \left( (\bar z-z_0)/l\right),
\label {eelastica}
\end{equation}
where $g(x)$ is 
a dimensionless geometrical function\cite{poisson}. 
We have determined $g$ numerically
and the result is shown in Fig. \ref{f2} as a continuous line.
We can rationalize the general form of the function $g(x)$,
considering how the elastic energy is distributed along the $z$
direction. 
The total energy $e_{\rm el}$ is the integral over $z$ of 
the density of elastic energy $\delta e_{\rm el}(z)$.
In  Fig. \ref{f1} we show qualitatively the form of $\delta e_{\rm el}(z)$
for the cases 
$\bar z\ll z_0$, and $\bar z\gg z_0$. 
Let us consider first the case $\bar z\ll z_0$ (Fig. \ref{f1}(a)).
$\delta e_{\rm el}(z)$ is zero for $z> z_0$, since as we already 
discussed, the material has to be unstressed for $z\rightarrow \infty$. 
In the region $z_0>z>\bar z$ there is a rather constant energy density, 
associated to the change in humidity concentration, which cannot be compensated
by a change in density of the material since the material here is attached to
the part below $z_0$\cite{nota2}.
Around the position of the
crack front there is an increase of the stored elastic energy, 
which is associated to the elastic energy around the
tips of the cracks. For $z<\bar z$ the elastic energy density 
goes to zero, since here the existence of the cracks 
has allowed to relieve the elastic energy accumulated 
previously to the cracks formation. 
The linear dependence of
$g(x)$ as $x\rightarrow -\infty$ (i.e. for $\bar z\ll z_0$)
comes from the energy stored between 
$\bar z$ and $z_0$.
Let us consider now the case when $\bar z\gg z_0$ 
(Fig. \ref{f1}(b)). The elastic energy
becomes independent of $\bar z$, as the crack tips are in a region of material
that is unstressed. The constant value of $g(x)$ as $x\rightarrow +\infty$
comes from the energy stored in the 
independent stripes around the position of the humidity front, 
which in this limit is well behind the crack front. When $\bar z\simeq z_0$
there is a smooth crossover between the two limiting regimes.

To determine the actual position of the crack front in a realistic situation 
we will rely on the energetic argument. 
At any time during the drying process, the crack front will be located
at the position that minimizes the total energy of the system.
Equation (\ref{eelastica}) gives the elastic energy of a set of cracks
that have penetrated down to the position $\bar z$. 
In order to get the total energy, $e_{\rm el}$ 
has to be added to the energy cost of creating the cracks. This 
part, when measured
per unit of horizontal length, will be called the crack energy $e_{\rm ck}$, and
it is simply given in terms of the specific energy fracture of the
material $\eta$ in the form

\begin{equation}
e_{ck}=\eta \bar z/l 
\label {ecracks}
\end{equation}
In order to determine the value of $\bar z$ at which the fracture front prefers
to be located, we have to minimize the total energy 
$e_{\rm tot}=e_{\rm el}+e_{\rm ck}$.
The result we obtain is shown in Fig. \ref{f3}, where we plot the most convenient
position of the crack front  
as a function of the parameter $u\equiv\eta /[C(\alpha \Delta h)^2 l]$ .
As we see, for very small value of $u$, $x\equiv (\bar z-z_0)/l$ 
takes large and positive
values, i.e., the crack front is located well below the humidity front. This
is due to the negligible contribution of the crack energy compared to the 
elastic energy in this case. As $u$ increases $x$ decreases, crossing zero
(namely, the crack front coinciding with the humidity front) at $u\simeq 0.66$.
$x$ tends to $-\infty$  as $u$ approaches the limiting value $u_0\simeq 1.14$. 
In fact, for $u>u_0$ the crack energy is so high that cracks do not penetrate
the sample at all\cite{vidrio}.

Under the conditions analyzed in this section, the crack front will be
located at a position such that $(\bar z-z_0)/l$ is given by the function plotted in
Fig. \ref{f3}. The crack front advances only due to changes in the position of the 
humidity front $z_0$, keeping always $\bar z-z_0$ as constant.

\section{Stability of the flat crack front}

The previous analysis has assumed that all cracks penetrate down to
a uniform depth $\bar z$, and has focused on what the value of $\bar z$ is, on
energetic grounds. It has to be complemented however with 
a stability analysis of the crack front. The necessity of this is clear
from the following example. If a material with an array of vertical cracks is 
loaded with a uniform horizontal stress (this situation can be
thought to be realized in our case
if $\bar z\ll z_0$), there will be 
typically a single crack that propagates and fractures the material. This
is a consequence of the fact that as soon as a single crack moves forward
a small distance, the stress at its tip increases, and those at the tips
of the other cracks decrease. This generates an unstable, rapid propagation
of a single crack. We will see that in our case, this can be compensated by the
fact that stresses decrease ahead of the humidity front, and this can 
stabilize a flat crack front.

The same kind of energetic arguments used in the previous section 
will be used to determine the stability conditions of the crack front.
Consider an evenly spaced set of cracks, labelled sequentially
by an index $j$, where now the tips of the cracks are at vertical positions $z^j$,
which can be slightly displaced from the mean position $\bar z$, i.e., 
$z^j=\bar z+\delta_ij$, with $\sum_j \delta_j=0$. The horizontal 
positions of the cracks are given by $x^j=jl$. The elastic energy of 
this configuration contains a term of the form (\ref{eelastica}), plus a correction
that can be expanded in powers of $\delta_j/l$. The first order term
in this expansion vanishes, as $\sum_j \delta_j=0$. The second order term can
be written in the form
\begin{equation}
\Delta e_{\rm el}=l^{-2}(a\sum_j \delta_j^2+b\sum_j \delta_j\delta_{j+1}
+c\sum_j \delta_j\delta_{j+2}+...)
\end{equation}
Successive term contain `interactions' between more distant cracks.
As the elastic energy is a local quantity, we expect $|a|<|b|<|c|<...~$.  
Any small displacement of the crack front, defined by the quantities
$\delta_j$ can be decomposed in a sum of `normal modes', by going
to the Fourier representation
\begin{equation}
\delta_j\equiv \int dk \tilde \delta_k \exp (ikx^j),
\end{equation}
and the energy decomposes into independent term, in the form
\begin{equation}
\Delta e_{\rm el}\sim \sum_k |\tilde \delta_k|^2 (a+b\cos (2\pi k/l)+c\cos (4\pi k/l)+...).
\end{equation}
In order for the flat crack front to be stable, $\Delta e_{\rm el}$ must be positive
for any choice of the $\delta _i$, and thus of the $\tilde \delta _k$.
This implies that 
\begin{equation}
(a+b\cos (2\pi k/l)+c\cos (4\pi k/l)+...)
\label{abc}
\end{equation}
must be positive for all $k$. In the limit of very small $k$, 
$\Delta e_{\rm el}$
is equivalent to a uniform advance of the crack front, and then is 
has to be a positive quantity, as the curvature of $e_{\rm el}(\bar z)$ is
always positive (see (\ref{eelastica}) and Fig. \ref{f2}).
On the other hand, It will be proportional to the 
sum of all coefficients in expression (\ref{abc}), i.e., 
$\Delta e_{\rm el}^{k\rightarrow 0}\sim a+b+c+...$.
Then it is clear that if an instability exists for some 
value of $k$, it will occur at $k=\pi/l$, where 
$\Delta e_{\rm el}^{k=\pi/l}\sim a-b+c-...~$.
Then we will analyze the stability of the flat crack front against 
a perturbation with $k=\pi/l$.
We took the equilibrium position of 
a crack front obtained in the previous
section (plotted in Fig. \ref{f3}) and calculated numerically 
the quadratic change in energy $\Delta e_{\rm el} = \chi(x) 
(\alpha \Delta h)^2 C  \delta ^2/l$ 
when a perturbation of $k=\pi/l$ and amplitude 
$\delta$ is introduced.
The dimensionless function $\chi (x)$ can be seen in Fig. \ref{f2} (dashed line).
We see that $\chi$ is positive 
(negative) for $x$ greater (lower) than $x_{\rm cr}\simeq -0.038$.
In this way we obtain that there are two regions in the plot of Fig. \ref{f3}.
The one at the left of $u_{\rm cr}\simeq 1.02$ corresponds to a stable situation. 
If the parameters 
of the system make the pattern of cracks to lie in that region, then the
time evolution of the drying process (i.e., the increase with time of $z_0$)
will produce a smooth advance of the crack front, keeping always the
same value of the distance $\bar z-z_0$ to the humidity front. 
At the right of $u_{\rm cr}$, the pattern is unstable. Should we have
one of those patterns at a given time, it will immediately propagate
forward some of its cracks (ideally, one of each two cracks), 
in order to reach a stable situation. 
This will imply in particular that some cracks will remain halted. The
further evolution of the crack front will correspond to a new crack pattern
with less cracks (i.e., with larger $l$) being propagated. 

\section{Appearance and ordering of cracks}

Up to now we have assumed that a set of evenly spaced cracks exists,
and we focused on its stability conditions. We will study now how this
pattern can appear, starting with the process at the surface.
Cracks are not expected to appear evenly spaced at the surface.
In fact, it is known
in one-dimensional models of surface fragmentation \cite{1d} (that can 
be used to represent
the first stage of cracking of our two-dimensional
problem) that the distribution of fragment length is strongly 
dependent on the presence of small inhomogeneities in the material, 
making the fragment length distribution broad. But we will see that
as the superficial cracks penetrate the sample they become evenly
spaced. 
We will first discuss how the process of nucleating new cracks at the
surface is, and then argue that the cracks become evenly spaced as they penetrate
the sample.

\subsection{The appearance of cracks at the surface}

When the drying front penetrates the material from its surface, and before
the material gets cracked, the stresses can be very simply 
calculated.
The state of the system corresponds to the material being completely unstrained 
horizontally
(considering the unstrained state as that corresponding to the humidity value
$h_0$ that occurs well inside the material). 
Under this condition a uniform horizontal
stress $T_0$ appears for all $z<z_0$, 
which is simply calculated as $T_0=C\alpha \Delta h$. 
In order for the first crack to appear, this value has to overcome 
the uniform traction resistance of the material $T_r$ \cite{microck}: 
\begin{equation}
C\alpha \Delta h>T_r. 
\label{tr}
\end{equation}
If this relation is not satisfied, no cracks will appear whatsoever. 
Assuming that the relation (\ref{tr}) is satisfied, 
the first crack will nucleate at the surface. This crack
penetrates the sample as long as this penetration reduces the total energy 
of the system. In our case, the crack penetrates only 
down to a distance $d_0$ of the order of $z_0$, where the humidity front 
is located%
. The horizontal stress at the surface is now zero at the position of the
crack, and increases as we move away from it, reaching the value $F_0$
at large distances of the crack. This means that new cracks will nucleate
away of the first one, in regions where relation (\ref{tr}) is still satisfied.
The number of cracks nucleated at the surface will be typically 
the minimum number that makes
the horizontal stress at every point of the surface
to be lower than $T_r$. The typical distance between cracks $l$
can then be estimated to be\cite{1d}
\begin{equation}
l\simeq d_0 f(T_0/T_r)
\end{equation}
where the geometrical function $f(T_0/T_r)$ goes to infinity when
$T_0\rightarrow T_r^+$, and is of order 1 for $T_0 \gg T_r$. 
Typically, $l$ is a few times $d_0$.

If we stick to the ideal sharp drying front we have 
been studying, the first crack appears for an infinitesimally
small value of $z_0$, the only restriction is that the relation
(\ref{tr}) is satisfied. Therefore, since the depth of the first crack $d_0$
is of the order of $z_0$, we should expect a very dense set 
of superficial cracks. However, this unrealistic situation is
removed if we note that any physical drying front will have some typical (finite) 
humidity gradient $\nabla h$. At the moment when relation (\ref{tr})
is first satisfied (with $\Delta h$ being now the difference between humidity 
concentrations right at the surface, and well inside the material), 
the first crack will penetrate down to a distance
\begin{equation}
d_0\simeq \Delta h/\nabla h=T_r/(C\alpha \nabla h),
\end{equation}
and the typical distance between cracks
at the surface will be a few times this distance. 

\subsection{How the distribution of cracks gets uniform}

The cracks that appear at the surface of the sample 
nucleate at positions that are strongly influenced
by the presence of small inhomogeneities in the sample \cite{1d}.
But a superficially uneven set of cracks has a tendency to become 
evenly spaced as it penetrates the sample.
The reason for this tendency can be understood on an 
energetic basis. For a fixed number of cracks the evenly spaced
configuration corresponds to the minimum of elastic energy.
It is then clear, using the energetic arguments, that this configuration
will be approached during the propagation process\cite{ronsin2}.

It is important to clarify the different effects 
that inhomogeneities have on the creation and propagation of cracks.
For the superficial layer, inhomogeneities are relevant, and responsible for the broad 
distribution of fragment lengths\cite{1d}, since at the beginning all the surface is 
uniformly stressed, and tiny differences in the properties of the material
will dictate which point of the surface will fail first.
However, once the array of cracks has been
defined at the surface, inhomogeneities play a secondary role, since new
cracks do not appear during penetration, and the elastic energy
is only very weakly dependent on the precise distribution
of defects. Then the evolution 
of the pattern is basically the same as if inhomogeneities
were absent.

The previous energetic arguments predict the trend towards evenly spaced cracks as
they penetrate the sample. 
Stress analysis allows to re-obtain and make this result
quantitative.
We have calculated the 
stresses that are present in the semi-infinite sample with a sharp
drying front, in the presence of a particular uneven set of parallel cracks. 
We took an infinite set of cracks separated sequentially by distances $l_1$ and $l_2$.

For this geometrical configuration we have determined 
numerically the direction of maximum
opening stress (mode I), at the tips of the fractures, and characterized it
by its angle $\theta$ with respect to the direction of forward advance. 
The first result is that $\theta$ always points in the direction of 
making the pattern more evenly spaced.
The actual value of $\theta$ is absolutely independent
of the values of $C$ and $\eta$ of the material, and also on the value $\Delta h$ of the 
step of the humidity front. It is in fact a quantity that only depends on the Poisson ratio
$\nu$ of the material and the values of $x\equiv (\bar z-z_0)/l$ and $(l_1-l_2)/(l_1+l_2)$. 
The results for the angle $\theta$, for a material with 
$\nu=1/3$ can be summarized as follows: $\theta$ is
rather independent of $x$, and within a maximum 10 \% error, it can be written
as 
\begin{equation}
\theta {\rm (deg)}\simeq 29 (l_1-l_2)/(l_1+l_2)\,
\label{ll}
\end{equation}
The value of $\theta$ vanishes for $l_1=l_2$, as in this case the pattern is actually
evenly spaced, and the cracks advance straightforwardly by symmetry. The limiting
case $l_1\gg l_2$ gives $\theta\simeq 29$ deg\cite{29}.
This is the maximum bending we can expect from a given crack, in the process of
uniformising the widths of the stripes. Note that as a consequence of the fact that
there are no typical lengths in (\ref{ll}),
the length needed for the pattern to become evenly spaced will
be proportional to the typical width of the stripes $l$.

Once the cracks have started to deviate in order to reduce the elastic energy as much as 
possible, the further detailed prediction of its evolution becomes more difficult, since now
we should calculate stresses ahead of a set of non parallel fractures (or alternatively,
the elastic energy of this nontrivial configurations of curved cracks). However,
the main conclusion that the pattern becomes evenly spaced in a distance of the order
of the stripes widths remains\cite{coupled}.

\section{Dependences on the characteristics of the drying front}

The sharp drying front is an idealization that is never exactly realized in
practice. For the experiments of Ref. \cite{canad}, for instance, a typical
distance over which humidity changes is expected. In other cases in which
the drying is through surface diffusion, the humidity profiles are still
smoother. 
As we have already discussed the stability reason of a flat crack front
in the case of a sharp humidity step is due to the rapid reduction of stresses
ahead of the crack front. Smoother humidity profiles will produce a weaker tendency
to generate a flat crack front, and it can even occur that a flat crack 
front is never stable
if the humidity profile is smooth enough. In this case a pattern of bifurcations
of the crack front has been predicted\cite{bb}.
Note that an additional stabilizing factor of the crack front exists when the drying
is favored by the presence of the cracks themselves\cite{bahr}, a case we have not addressed here.

\section{Implications for the three dimensional case}

The same phenomenon we have been discussing acquires novel characteristics in a three 
dimensional geometry. It has a beautiful realization in the geological 
formations named columnar basalts, which can be reproduced in a kitchen experiment
using corn-starch\cite{muler}. When an originally hot volcanic lava flow starts
to cool down (after solidification), the thermal stresses generate cracks at the
surface, which propagate progressively towards the interior of the igneous body.
There are many coincidences with the two dimensional case, and also some new features
\cite{basalts,jr0,jr1}.
Cracks are known to appear at the surface and at later times propagate to the interior.
At the surface the pattern of cracks is rather disordered, but becomes progressively
ordered as it penetrates the material. In three dimensions, the tendency to order
manifest in progressive tendency to form a polygonal pattern of cracks, in planes parallel
to the surface. Qualitatively this tendency to order can be understood on the same basis
as the two dimensional case, as a tendency of the system to reduce its total energy as
much as possible during the evolution. 
It poses however some interesting problems, because typically,
the expected perfect hexagonal pattern is not reached, but a collection of polygons
with different number of sides and areas. It has been recently demonstrated\cite{jr1}
that this is a consequence of the fact that the patterns starts being disordered
at the surface, and in the process of minimizing the energy it is not able to
reach the absolute minimum (namely, the honeycomb pattern) but is trapped in
a metastable minimum.
Based in the arguments for the two dimensional case, we can expect also in three dimensions
that the typical width of the columns is set by 
the typical distance between cracks at the surface,
and it is not modified with the further penetration. In fact this is what is
experimentally observed,
since columns are seen to keep its horizontal size over distances that reach
hundred times the width of the columns.
For the three dimensional case, the cooling through diffusion plus convection 
(and/or radiation)
at the surface is the most realistic model of cooling to consider. Although
this type of cooling can lead in two dimensions to the instability of the flat
crack front, this is not necessarily so in three dimensions, since in this case
all fractures form a connected
structure, and this generates an additional tendency to stabilize the flat crack front.
These issues and some additional ones particular of the 3D case will be discussed in 
a forthcoming publication.

\section{Summary and conclusions}

In this paper we have focused on a two dimensional material that cools
down or dessicates from one edge. We studied the conditions under which 
a set of cracks penetrate the sample during the process. We showed that
these cracks have a tendency to become evenly spaced, an analyze the conditions
for which all the tips of the cracks form a plane front. Our results
provide a quantitative framework to analyze recent experiments.
Extensions to the three dimensional case are expected to provide both a 
qualitatively similar scenario, and also some new features that will be fully 
developed elsewhere.

\section{Acknowledgments}

This work was financially supported by CONICET (Argentina). Partial support
from Fundaci\'on Antorchas is also acknowledged.

\begin{figure}
\narrowtext
\epsfxsize=3.3truein
\vbox{\hskip 0.05truein
\epsffile{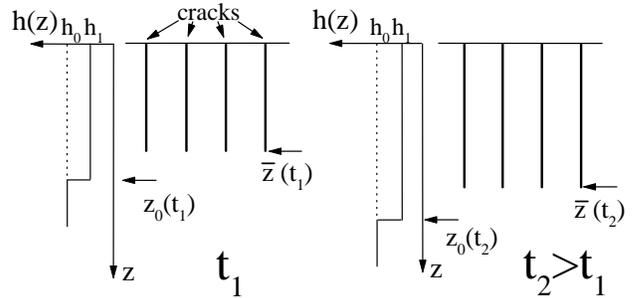}}
\caption{Sketch of the process studied. There is a sharp drying profile at
depth $z_0$, with $z_0$ increasing slowly with time. As this occurs, the crack
front (located at $\bar z$)
penetrates more deeply into the material. The distance $\bar z-z_0$ (which can be
positive or negative, depending on the parameters) remains constant in time.}
\label{f0}
\end{figure}

\begin{figure}
\epsfxsize=3.3truein
\vbox{\hskip 0.05truein
\epsffile{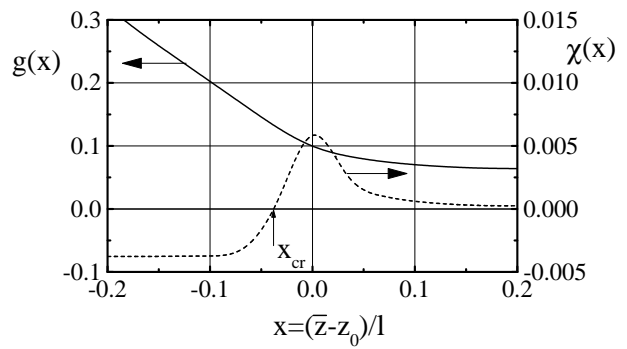}}
\caption{The functions $g(x)$ (continuous line) and $\chi (x)$ (dotted line)
for a material with Poisson ratio $\nu=1/3$. 
The values of $x$ for which $\chi (x)>0$
are those corresponding to a stable flat crack front.
} 
\label{f2}
\end{figure}

\begin{figure}
\narrowtext
\epsfxsize=3.3truein
\vbox{\hskip 0.05truein
\epsffile{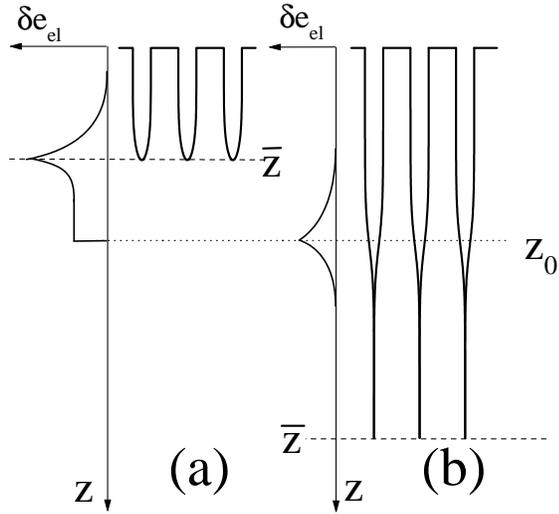}}
\caption{Sketch of the density of energy $\delta e_{\rm el}$ as a function of $z$
for a set of fractures
that has penetrated down to $\bar z<z_0$ (a), and $\bar z>z_0$ (b). The contraction
of the stripes is exaggerated for clarity.}
\label{f1}
\end{figure}

\begin{figure}
\narrowtext
\epsfxsize=3.3truein
\vbox{\hskip 0.05truein
\epsffile{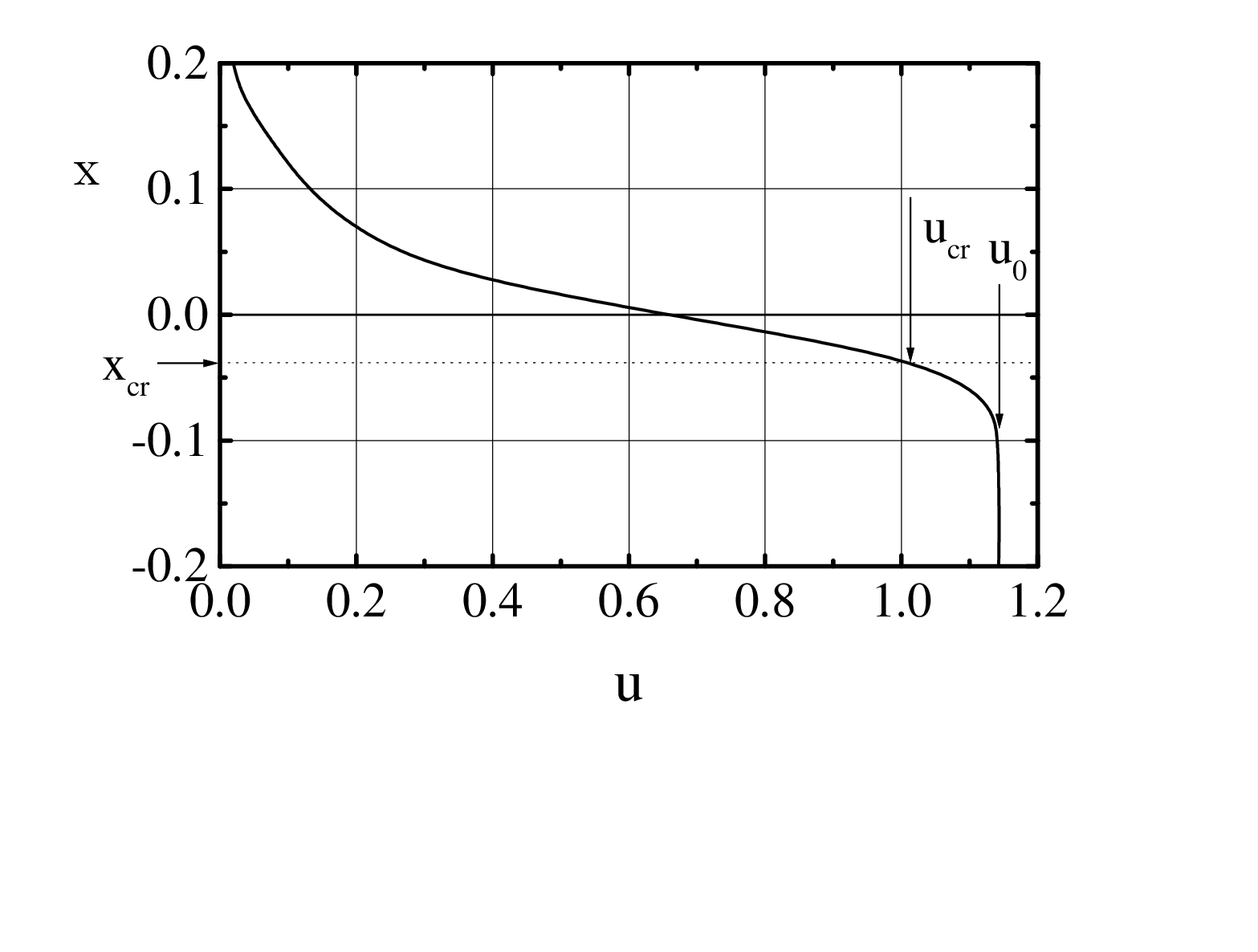}}
\caption{Position of the crack front [$x\equiv (\bar z-z_0)/l$] as a function of
$u\equiv \eta/[C(\alpha \Delta h)^2l]$, obtained by minimizing the
total energy of the system. The region at the left of $u_{\rm cr}$ 
corresponds to a stable situation in which all tips of the cracks form a stable
crack front, whereas at the right of this value the flat crack front is
unstable.}
\label{f3}
\end{figure}


\begin{references}
\bibitem{nota10}We are implicitly assuming here that the minimum in which 
the system is located does not destabilizes upon a change of the external
conditions.
\bibitem{elasticidad}To be more precise, we are assuming
that the size of the plastic region near the tips of the cracks
is always small compared to the distance between cracks. The
energy dissipated in the plastic region is accounted for in the
numerical value of the fracture energy $\eta$ defined below.
\bibitem{zz}We are assuming here that $|\bar z-z_0|\ll z_0$, in such
a way that surface effects are negligible.
\bibitem{poisson}To simplify the presentation we will consider a material
with a fixed Poisson ratio $\nu$. In a general case, all constants and functions
introduced in the analysis must be considered to be also functions of $\nu$.
\bibitem{nota2} The material is vertically compressed in this region in such a 
way that vertical stress vanishes. However, it is the
impossibility of producing a horizontal compression what generates a horizontal
stress and then the finite value of this part of the 
elastic energy.
\bibitem{vidrio} From the results of Fig. \ref{f3} we can obtain the result that
thiner stripes will have their crack front more retracted compared
to thicker stripes (since $u\sim l^{-1}$), and that a minimum $l$ is necessary
for propagation. This is qualitatively
comparable to experimental results \cite{ronsin} 
showing that a single crack propagating in a glass plate of width $l$ between two
thermal baths has an equilibrium position that retracts as $l$ reduces, and also 
that the propagation is halted if $l$ is lower than some critical value.
\bibitem{microck}This is the traction resistance value for the bulk material, 
which of course takes into account the existence of defects (such as microcracks)
in the material.
\bibitem{ronsin2}The tendency to uniformise the stripe width $l$ is also consistent
with the experimentally observed stability of a crack propagating exactly at the middle
of a glass plate, driven by a temperature difference (see Ref. \cite{ronsin}).
\bibitem{coupled}During the process of uniformising an originally disordered
set of cracks, the crack front can deviate slightly from the sharp
form expected for evenly spaced cracks.
For instance, for a set of fractures with a typical width
of columns $l$ that depends smoothly on the horizontal coordinate, the fracture
front will be forwardly advanced at those position with largest $l$, and retracted
at positions with smallest $l$.
\end{references}
\end{document}